\documentclass[12pt]{article}

\newcommand{\be}{\begin{equation}}
\newcommand{\ee}{\end{equation}}
\newcommand{\bi}[1]{\vspace{-3mm} \bibitem{#1}}

\usepackage{epsfig,amsmath,amssymb,graphics,graphicx} 

\begin{document}
\begin{center}

{\it Comptes Rendus Mecanique. 
Vol.343. No.1. (2015)  57-73.}
\vskip 3mm

{\bf \large Elasticity of Fractal Material by 
Continuum Model with Non-Integer Dimensional Space} \\

\vskip 7mm
{\bf \large Vasily E. Tarasov} \\
\vskip 3mm

{\it Skobeltsyn Institute of Nuclear Physics,\\ 
Lomonosov Moscow State University, Moscow 119991, Russia} \\
{E-mail: tarasov@theory.sinp.msu.ru} \\

\begin{abstract}
Using a generalization of vector calculus for space with 
non-integer dimension, 
we consider elastic properties of fractal materials. 
Fractal materials are described by continuum models
with non-integer dimensional space. 
A generalization of elasticity equations 
for non-integer dimensional space, 
and its solutions for equilibrium case of 
fractal materials are suggested.
Elasticity problems 
for fractal hollow ball and cylindrical fractal elastic pipe 
with inside and outside pressures, 
for rotating cylindrical fractal pipe,
for gradient elasticity and thermoelasticity 
of fractal materials are solved.
\end{abstract}

\end{center}


\section{Introduction}

Fractals are measurable metric sets with non-integer 
Hausdorff dimensions \cite{Fractal1,Fractal2}. 
The main characteristic of fractal set is non-integer 
Hausdorff dimension that should be observed on all scales.
The Hausdorff dimension is a local property, i.e. this dimension characterize 
(measure) property of a set of distributed points in the limit 
of a vanishing diameter, which is used to cover subset of the points.
By definition the Hausdorff dimension  
requires the diameter of the covering sets to vanish.
In general, real materials have a characteristic smallest length scale $R_0$
such as the radius of a particle such as atom or molecule.
In fractal materials the fractal structure cannot be observed on all
scales but only those for which $R > R_0$, where
$R_0$ is the characteristic scale of the particles.
For real materials, a non-integer mass dimension can be 
used instead of Hausdorff dimension.
The mass dimension described how the mass of a medium region 
scales with the size of this region, 
where we assume unchanged density.
For many cases, we have an asymptotic relation between
the mass $M(W)$ of a ball region $W$ of material, 
and the radius $R$ of this ball. 
The mass of fractal material  
satisfies a power-law relation $M(W) \sim R^{D}$.
The parameter $D$ is called the non-integer mass dimension
of fractal material. 
This parameter does not depend on the shape 
of the region $W$, or on whether the packing of sphere 
of radius $R_0$ is close packing, a random packing 
or a porous packing with a uniform distribution of holes.
Therefore fractal material can be considered as 
a medium with non-integer mass dimension. 
Although, the non-integer dimension does not reflect 
completely the geometrical and dynamical properties of 
the fractal materials, 
it nevertheless permits a number of important conclusions 
about the behavior of materials. 
It allows us to use effective models that take into 
account non-integer dimensions.

We can distinguish the following approaches to 
formulate models of fractal materials:

1) Approach based on methods of "Analysis on fractals" 
\cite{Kugami,Strichartz-1,Strichartz-2,Harrison,Kumagai,DGV}
can be considered as the most rigorous approach to 
describe fractal materials. 
Unfortunately a possibility of application of the 
"Analysis on fractals" to solve real problems of fractal material
now is very limited due to weak development of this area of mathematics. 

2) To describe fractal material we can apply special continuum models suggested 
in \cite{PLA2005-1,AP2005-2,IJMPB2005-2,MPLB2005-1} 
and then developed in the works  
\cite{MOS-1}-\cite{MOS-5b} and \cite{TarasovSpringer}.
These models can be called 
the fractional-integral continuum models.
In this approach we use integrations of non-integer orders, and  
two different notions such as density of states and 
distribution function \cite{TarasovSpringer}.
The order of the fractional integrals is equal to 
mass dimension of fractal materials.
The kernels of these integrals are defined by the power-law type
of density of states. 

3) Fractional derivatives of non-integer orders
are used to describe some properties of fractal materials. This approach has been suggested 
in papers \cite{CCC2001,CCC2002,CCC2003},
where so-called local fractional derivatives are used, 
and then developed in the works 
\cite{CCC2004a,CCC2004b,CCC2004c,CCC2004d,CCC2004e,CCSPZ2009,CCC2009}.
These models can be called the fractional-differential models.

4) Fractal materials can be described by using
the theory of integration and differentiation 
for non-integer dimensional spaces 
\cite{Collins,Stillinger,PS2004}.
Fractal materials are described as continuum
in non-integer dimensional spaces.
The dimension of the spaces are equal to the mass dimensions of fractal materials.


Unfortunately there are not enough differential equations 
that are solved for various problems 
for fractal materials 
in the framework of the fractional-differential model
and by methods of "Analysis on fractals". 

The fractional-integral continuum models 
are used to solve differential equations 
for various problems of
elasticity of fractal materials 
\cite{MOS-3,MOS-4,MOS-4b,MOS-4c,MOS-5,MOS-5b}, and
thermoelasticity of fractal materials 
\cite{MOS-TE1,MOS-TE2}.

Continuum models with non-integer dimensional spaces 
are not currently used to describe elasticity of 
fractal materials.
In this paper, we consider approach based on
the non-integer dimensional space
to describe elasticity of isotropic fractal materials.
The main difference of 
the continuum models with non-integer dimensional spaces
and fractional-integral continuum models suggested in 
\cite{PLA2005-1,AP2005-2,IJMPB2005-2,MPLB2005-1,TarasovSpringer}
may be reduced to the following:
(a) Arbitrariness in the choice of the numerical factor
in the density of states is fixed by the equation
of the volume of non-integer dimensional ball region.
(b) In the fractional-integral continuum models 
the differentiations are integer orders whereas
the integrations are non-integer orders.
In the continuum models with non-integer dimensional spaces
the integrations and differentiations are defined for
the spaces with non-integer dimensions.
The power law $M \sim R^{D}$ can be naturally 
derived by using the integrations 
in non-integer dimensional space \cite{Collins},
where the dimension of this space is equal to
the mass dimension of fractal material.

A vector calculus for non-integer dimensional space proposed in this paper 
allows us to use continuum models, which are based 
on non-integer dimensional space, to describe fractal materials.
This is due to the fact that
although the non-integer dimension  
does not reflect all geometrical and dynamical
properties of the fractal materials, it nevertheless allows us
to get important results about the behavior
of fractal materials.
Therefore continuum models with non-integer dimensional spaces
can describe a wide class of fractal materials.

Integration over non-integer dimensional spaces 
are actively used in the theory of critical phenomena and 
phase transitions in statistical physics 
\cite{WilsonFisher,WK1974}, and 
in the dimensional regularization of
ultraviolet divergences 
in quantum field theory \cite{HV1972,Leibbrandt,Collins}. 
The axioms for integrations in non-integer dimensional space are proposed in \cite{Wilson,Stillinger} and this type of integration is considered in the book by Collins \cite{Collins}
for rotationally covariant functions. 
In the paper \cite{Stillinger} a mathematical basis of integration on non-integer dimensional space is given,
and a generalization of the Laplace operator 
for non-integer dimensional spaces is suggested. 
Using a product measure approach,  
the Stillinger's methods \cite{Stillinger}
has been generalized by Palmer and Stavrinou \cite{PS2004} 
for multiple variables case
with different degrees of confinement in orthogonal directions. 
The scalar Laplace operators suggested 
by Stillinger in \cite{Stillinger} 
and by Palmer, Stavrinou in \cite{PS2004}
for non-integer dimensional spaces, 
have successfully been used for 
effective descriptions in physics and mechanics. 
The Stillinger's form of Laplacian 
for the Schr\"odinger equation in non-integer dimensional space 
is used by He \cite{XFHe1,XFHe2,XFHe3} to describe
a measure of the anisotropy and 
confinement by the effective non-integer dimensions. 
Quantum mechanical models with non-integer (fractional) 
dimensional space has been discussed in 
\cite{Stillinger,PS2004,Thilagam1997b,MA2001a,MA2001b,QM1,QM5}
and \cite{Muslih2010,MA2012,QM8,QM9}.
Recent progress in non-integer dimensional space approach
also includes description of
the fractional diffusion processes in
non-integer dimensional space in \cite{LSTLRL}, and
the electromagnetic fields in non-integer dimensional space
in \cite{MB2007,BGG2010,MSBR2010}
and \cite{ZMN2010,ZMN2011a,ZMN2011b,ZMN2011c}. 


Unfortunately, in the articles \cite{Stillinger,PS2004} 
are proposed only the second order differential operators 
for scalar fields in the form of the scalar Laplacian 
for the non-integer dimensional space. 
A generalization of the vector Laplacian \cite{VLap}
for the non-integer dimensional space is not suggested. 
The first order operators such as gradient, 
divergence, curl operators, and the vector Laplacian 
are not considered in \cite{Stillinger,PS2004} also.
In the work \cite{ZMN} 
the gradient, divergence, and curl operators
are suggested only as approximations of the square of 
the Laplace operator.
Consideration only
the scalar Laplacian in non-integer dimensional space approach
greatly restricts us in application of continuum models
with non-integer dimensional space for fractal materials and material.
For example, we cannot use the Stillinger's form of Laplacian
for the displacement vector field ${\bf u}({\bf r},t)$ in 
theory of elasticity and thermoelasticity of fractal materials, 
for the velocity vector field ${\bf v}({\bf r},t)$
in hydrodynamics of fractal fluids, 
for electric  and magnetic vector fields  
in electrodynamics of fractal media
in the framework non-integer dimensional space approach.

In this paper, we define
the first and second orders differential vector
operations such as gradient, divergence, 
the scalar and vector Laplace operators 
for non-integer dimensional space.
In order to derive the vector differential operators 
in non-integer dimensional space
we use the method of analytic continuation in dimension.
For simplification we consider rotationally covariant 
scalar and vector functions that are independent of angles.
It allows us to reduce differential equations 
in non-integer dimensional space to
ordinary differential equations with respect to $r$. 
The proposed operators  
allows us to describe fractal materials
to describe processes in the framework of continuum models
with non-integer dimensional spaces. 
In this paper we solve elasticity problems 
for fractal hollow ball with inside and  outside pressures,
for cylindrical fractal elastic pipe 
with inside and outside pressures, and
for rotating cylindrical fractal pipe,
for gradient elasticity and thermoelasticity 
of fractal materials.

\section{Differential and integral operators 
in non-integer dimensional space}

To derive equations for vector differential operators 
in non-integer dimensional space,
we use equations for the differential operators 
in the spherical (and cylindrical) coordinates in $\mathbb{R}^n$ 
for arbitrary $n$ to highlight the explicit 
relations with dimension $n$.
Then the vector differential operators  
for non-integer dimension $D$ 
can be defined by continuation in 
dimension from integer $n$ to non-integer $D$. 
To simplify we will consider only scalar 
fields $\varphi$ and vector fields ${\bf u}$ 
that are independent of angles 
\[ \varphi({\bf r}) =\varphi(r) ,
\quad {\bf u}({\bf r})={\bf u} (r) = u_r\, {\bf e}_r , \]
where $r=|{\bf r}|$ is the radial distance,
${\bf e}_r={\bf r} /r$ is the local orthogonal unit 
vector in the directions of increasing $r$,  
and $u_r=u_r(r)$ is the radial component of ${\bf u}$.
We will work with rotationally covariant functions only. 
This simplification is analogous to the simplification
for definition of integration over non-integer dimensional space
described in Section 4 of the book \cite{Collins}.

\subsection{Vector differential operators 
for spherical and cylindrical cases}


Explicit definitions of differential operators
for non-integer dimensional space can be obtained
by using continuation from integer $n$ 
to arbitrary non-integer $D$.
We mote that the same expressions can be obtained by using 
the integration in non-integer dimensional space
and the correspondent Gauss's theorem.

We define the differential vector operations
such as  gradient, divergence, the scalar and 
vector Laplacian for non-integer dimensional space. 
For simplifications, we assume that
the vector field ${\bf u}={\bf u}({\bf r})$ 
be radially directed and the scalar and vector 
fields $\varphi({\bf r})$, ${\bf u}({\bf r})$ 
are not dependent on the angles.

The divergence in non-integer dimensional space 
for the vector field ${\bf u}={\bf u}(r)$ is
\be \label{Div-D}
\operatorname{Div}^{D}_{r} {\bf u} = 
\frac{\partial u_r}{\partial r} + \frac{D-1}{r} \, u_r.
\ee

The gradient in non-integer dimensional space 
for the scalar field $\varphi=\varphi (r)$ is
\be \label{Grad-D}
\operatorname{Grad}^{D}_r \varphi = \frac{\partial \varphi}{\partial r} \, {\bf e}_r .
\ee

The scalar Laplacian in non-integer dimensional space 
for the scalar field $\varphi=\varphi (r)$ is
\be \label{S-Delta-D}
^S\Delta^{D}_r \varphi= 
\operatorname{Div}^{D}_r \operatorname{Grad}^{D}_{r} \varphi =
\frac{\partial^2 \varphi}{\partial r^2} + \frac{D-1}{r} \, 
\frac{\partial \varphi}{\partial r} .
\ee

The vector Laplacian in non-integer dimensional space 
for the vector field ${\bf u}=u_r(r) \, {\bf e}_r$ is
\be \label{V-Delta-D}
^V\Delta^{D}_r {\bf u} = 
\operatorname{Grad}^{D}_r \operatorname{Div}^{D}_{r} {\bf u} =
\Bigl(
\frac{\partial^2 u_r}{\partial r^2} + \frac{D-1}{r} \, 
\frac{\partial u_r}{\partial r}  -  \frac{D-1}{r^2} \, u_r
\Bigr) \, {\bf e}_r.
\ee

If $D=n$, equations (\ref{Div-D}-\ref{V-Delta-D})
give the well-known formulas for
integer dimensional space $\mathbb{R}^n$.


We can consider materials with axial symmetry, 
where the fields 
$\varphi(r)$ and ${\bf u}(r)=u_r(r) \, {\bf e}_r$
are also axially symmetric.
Let the $Z$-axis be directed along the axis of symmetry. 
Therefore we use a cylindrical coordinate system.

The divergence in non-integer dimensional space 
for the vector field ${\bf u}={\bf u}(r)$ is
\be \label{Div-DC}
\operatorname{Div}^{D}_{r} {\bf u} = 
\frac{\partial u_r}{\partial r} + \frac{D-2}{r} \, u_r.
\ee

The gradient in non-integer dimensional space 
for the scalar field $\varphi=\varphi (r)$ is
\be \label{Grad-DC}
\operatorname{Grad}^{D}_r \varphi = 
\frac{\partial \varphi}{\partial r} \, {\bf e}_r .
\ee

The scalar Laplacian in non-integer dimensional space 
for the scalar field $\varphi=\varphi (r)$ is
\be \label{S-Delta-DC}
^S\Delta^{D}_r \varphi = 
\frac{\partial^2 \varphi}{\partial r^2} + \frac{D-2}{r} \, 
\frac{\partial \varphi}{\partial r} .
\ee

The vector Laplacian in non-integer dimensional space 
for the vector field ${\bf u}=v(r) \, {\bf e}_r$ is
\be \label{V-Delta-DC}
^V\Delta^{D}_r {\bf u} = 
\Bigl(
\frac{\partial^2 u_r}{\partial r^2} + \frac{D-2}{r} \, 
\frac{\partial u_r}{\partial r}  -  \frac{D-2}{r^2} \, u_r
\Bigr) \, {\bf e}_r .
\ee


Equations (\ref{Div-DC}-\ref{V-Delta-DC}) can be
easy generalized for the case $\varphi=\varphi(r,z)$ and 
${\bf u}(r,z)=u_r(r,z) \, {\bf e}_r+ u_r(r,z) \, {\bf e}_z$.
In this case the curl operator for ${\bf u}(r,z)$
is different from zero, and
\be \label{Curl-DC}
\operatorname{Curl}^{D}_r {\bf u} = 
\left( \frac{\partial u_r}{\partial z} -
\frac{\partial v_z}{\partial r} \right) \, {\bf e}_{\theta} .
\ee


Equations (\ref{Div-D}) - (\ref{V-Delta-DC}) with $D=3$ and 
(\ref{Curl-DC}) give the well-known
expressions for the gradient, divergence, curl operator,
scalar and vector Laplacian operators

The proposed operators for $0<D<3$ allows us to reduce 
non-integer dimensional vector differentiations 
(\ref{Div-D}) - (\ref{V-Delta-D}) and 
(\ref{Div-DC}) - (\ref{V-Delta-DC})
to derivatives with respect to $r=|{\bf r}|$. 
It allows us to reduce partial differential equations for
fields in non-integer dimensional space to
ordinary differential equations with respect to $r$.


For a function $\varphi=\varphi(r,\theta)$ of radial distance $r$ 
and related angle $\theta$ measured relative 
to an axis passing through the origin, 
the scalar Laplacian in a non-integer dimensional space
proposed by Stillinger \cite{Stillinger} is
\be \label{NI-1}
^{St}\Delta^{D} = \frac{1}{r^{D-1}} \frac{\partial}{\partial r} \left( r^{D-1} \,\frac{\partial}{\partial r} \right) +
\frac{1}{r^2 \, \sin^{D-2} \theta} 
\frac{\partial}{\partial \theta} 
\left(  \sin^{D-2} \theta 
\frac{\partial}{\partial \theta} \right) ,
\ee
where $D$ is the dimension of space ($0<D <3$),
and the variables $r \ge 0$, $0\le \theta \le \pi$.
Note that
$(\, ^{St}\Delta^{D} )^2 \ne \, ^{St}\Delta^{2D}$.
If the function depends on radial distance $r$ 
only ($\varphi=\varphi(r)$), then
\be \label{NI-R}
^{St}\Delta^{D} \varphi (r) = \frac{1}{r^{D-1}} \frac{\partial}{\partial r} \left( r^{D-1} \,\frac{\partial \varphi (r)}{\partial r} \right) =
\frac{\partial^2 \varphi (r)}{\partial r^2} +
\frac{D-1}{r} \, \frac{\partial \varphi (r)}{\partial r} .
\ee
It is easy to see that 
the Stillinger's form of Laplacian $\, ^{St}\Delta^{D}$
for radial scalar functions $\varphi ({\bf r})=\varphi (r)$
coincides with the scalar Laplacian 
$\, ^S\Delta^{D}_r$ defined by (\ref{S-Delta-D}), i.e.,
\be \label{NI-R-2}
^{St}\Delta^{D} \varphi (r) = \, ^S\Delta^{D} \varphi (r) .
\ee
The Stillinger's Laplacian can be applied 
for scalar fields only. It cannot be used 
to describe vector fields ${\bf u}=u_r(r) \, {\bf e}_r$ 
because this Laplacian for $D=3$ is not equal to 
the usual vector Laplacian for $\mathbb{R}^3$,
\be
^{St}\Delta^{3} {\bf u}(r) \ne \, \Delta {\bf u}(r) =
\Bigl( 
\frac{\partial^2 u_r}{\partial r^2} + \frac{2}{r} \, 
\frac{\partial u_r}{\partial r} - \frac{2}{r^2} \, u_r
\Bigr) \, {\bf e}_r  .
\ee
The gradient, divergence, curl operator and vector Laplacian are not considered by Stillinger in paper \cite{Stillinger}.

\subsection{Integration over non-integer dimensional space}

Integration for non-integer values of dimension $D$ 
is defined by continuation in $D$ \cite{Leibbrandt,Collins}.   
The following properties suggested in \cite{Wilson}
for integrals in $D$-dimensional space   
are necessary for applications \cite{Collins}:

a) Linearity: 
\be \label{Ax-1}
\int \Bigl( af_1({\bf r})+bf_2({\bf r}) \Bigr) \, d^D {\bf r} =
a \int f_1({\bf r}) \, d^D r + b \int f_2({\bf r}) \, d^D {\bf r} ,
\ee
where $a$ and $b$ are arbitrary real numbers.

b) Translational invariance:
\be \label{Ax-2}
\int f({\bf r}+{\bf r}_0) \, d^D {\bf r} = 
\int f({\bf r}) \, d^D {\bf r} 
\ee
for any vector ${\bf r}_0$.

c) Scaling property: 
\be \label{Ax-3}
\int f(\lambda {\bf r} ) \, d^D {\bf r} = 
\lambda^{-D} \int f({\bf r}) \, d^D {\bf r}
\ee
for any positive $\lambda$.

Note that linearity is true of any integration, 
while translation and rotation invariance 
are basic properties of a Euclidean space.
The scaling property embodies the $D$-dimensionality. 
Not only the above three axioms are necessary, 
but they also ensure that integration is unique, 
aside from an overall normalization \cite{Wilson}. 
These properties must be used in order to have 
non-integer dimensional integrations \cite{Collins}.
These properties are natural  
in application of dimensional regularization
to quantum field theory \cite{Collins}.


In general, we can consider any functions 
of the components of its vector argument ${\bf r}$. 
However, we do not know the meaning 
of the components of a vector in non-integer dimensions. 
In this paper, we will work 
with rotationally covariant functions for simplification. 
So we will assume that $f$ is a scalar or vector function 
only of scalar products of vectors or of length of vectors. 
For example, in the elasticity theory, we consider the case,
where the displacement vector ${\bf u} ({\bf r})$, 
is independent of the angles
${\bf u} ({\bf r}) = {\bf u}(r)$, where $r=|{\bf r}|$.
The integration defined by equation (\ref{1-dimregint}) 
satisfies the properties (\ref{Ax-1}) - (\ref{Ax-3}). 


The non-integer dimensional integration for
scalar functions $f({\bf r})=f(|{\bf r}|)$ can be defined 
in terms of ordinary integration by the equation
\be \label{1-dimregint}
\int d^D {\bf r} \ f({\bf r}) 
= \int_{\Omega_{D-1}} d\Omega_{D-1} 
\int^{\infty}_0 dr \; r^{D-1} \; f(r) ,
\ee
where 
\be \label{SD-1}
\int_{\Omega_{D-1}} d\Omega_{D-1} = 
\frac{2 \pi^{D/2}}{\Gamma(D/2)} =S_{D-1}. \ee
Equation (\ref{SD-1}) with integer  $D=n$ 
gives the well-known area $S_{n-1}$ 
of $(n-1)$-sphere with unit radius.

As a result, we have \cite{Collins} 
the explicit definition of the continuation of integration
from integer $n$ to arbitrary fractional $D$ in the form
\be \label{1-dim-reg} 
\int d^D {\bf r} \ f(|{\bf r}|) = \frac{2 \pi^{D/2}}{\Gamma(D/2)} 
\int^{\infty}_0 dr \; r^{D-1} \; f(r) . \ee
This equation reduced non-integer dimensional integration 
to ordinary integration.
Therefore the linearity and translation invariance 
follow from linearity and translation invariance 
of ordinary integration. 
The scaling and rotation covariance are explicit properties of the definition.


As an example of applications of equation (\ref{1-dim-reg}),
we can consider non-integer dimensional integration for
the function
\be \label{f-example}
f({\bf r}^2)= \frac{{\bf r}^2+a}{{\bf r}^2+b} , \ee
where $a$ and $b$ are real numbers. 
The integral for (\ref{f-example}) can be explicitly computed
\be
\int d^D {\bf r} \, \frac{r^2+a}{r^2+b} =
(\pi b)^{D/2} \, (a/b-1) \, \Gamma(1-D/2) .
\ee
The other example is the non-integer dimensional integration is
\be
\int d^D {\bf r} \, \frac{r^{2 \alpha}}{(r^2+a^2)^{\beta}}= 
\frac{\Gamma(\alpha+D/2) \, 
\Gamma(\beta-\alpha-D/2)}{\Gamma(D/2) \Gamma(\beta)} \,
\pi^{D/2} \, a^{D+2\alpha -2\beta} ,
\ee
where $r=|{\bf r}|$.


\section{Mass and moment of inertia for fractal materials}

\subsection{Mass of fractal materials}

Fractal materials can be characterized by 
the relation between
the mass $M(W)$ of a region $W$ of fractal material,
and the size $R$ of the region containing this mass:
\be \label{MDW}
M_D(W) =M_0 \ \left(\frac{R}{R_0}\right)^D , 
\quad R/R_0 \gg 1 . \ee
The parameter $D$ is called the non-integer mass dimension
of fractal material. 
The parameter $D$, does not depend on the shape 
of the region $W$, or on whether the packing of sphere 
of radius $R_0$ is close packing, a random packing 
or a porous packing with a uniform distribution of holes.
The cornerstone of fractal materials is the non-integer mass dimension.
The mass dimension of real fractal materials
can be measured by box-counting method, 
which means drawing a box of size $R$ 
and counting the mass inside.

The fractality of material means than the mass of 
the region $W \subset \mathbb{R}^3$ 
increases more slowly than the 3-dimensional volume of this region.
For the ball region of the fractal medium, 
this property can be described by the power law $M_D(W) \sim R^{D}$, 
where $R$ is the radius of the ball. 

Fractal material is called homogeneous 
if the power law $M_D(W) \sim R^{D}$ does not depend on 
the translation of the region. 
The homogeneity property of the material
can be formulated in the form:
For all two regions $W_1$ and $W_2$ of the homogeneous fractal material 
with the equal volumes $V_D(W_1)=V_D(W_2)$, 
the masses of these regions are equal $M_D(W_1)=M_D(W_2)$. 
 
The power law (\ref{MDW}) can be naturally 
derived by using the integration in the non-integer dimensional space such that the space dimensions is equal to 
the mass dimension of the material. 

The mass of the region $W$ of fractal material in $W$ 
can be calculated by the integral 
in non-integer dimensional space
\be \label{1-MW3} 
M_D(W) = \int_{W} \rho({\bf r}) \, d^D {\bf r} ,
\ee
where ${\bf r}$ is dimensionless vector variable. 
For a ball region $W$ with radius $R$ and density 
$\rho({\bf r})=\rho_0 =\operatorname{const}$, we get 
the mass is defined by
\be \label{M-D}
M_D(W) = \rho_0 \, V_D =
\frac{\pi^{D/2} \, \rho_0}{\Gamma(D/2+1)} \, R^{D} .
\ee
This equation define the mass of the 
fractal homogeneous ball. 
For $D=3$, equation (\ref{M-D}) gives the well-known 
equation for mass of non-fractal ball
$M_3 =(4 \rho_0 \pi)/3 R^3$ because 
$\Gamma(3/2) =\sqrt{\pi}/2$ and $\Gamma(z+1)=z \, \Gamma(z)$.

\subsection{Moment of inertia of fractal materials}

Let us consider a calculation of scalar moment of inertia $I(t)$, which is used when the axis of rotation is known. 
The scalar moment of inertia of a rigid body with density 
$\rho^{\prime}({\bf r}^{\prime},t)$ 
with respect to a given axis is defined by the volume integral
\be \label{1-(1)}
I^{\prime}(t)=\int_W \rho^{\prime}({\bf r}^{\prime},t) \; 
{\bf r}^{\prime \,2 }_{\perp} \; d^3 {\bf r}^{\prime},
\ee 
where $({\bf r}^{\prime})^2_{\perp}$ is 
the perpendicular distance from the axis of rotation, and 
$dV^{\prime}_3=dx^{\prime}_1 dx^{\prime}_2 dx^{\prime}_3$.  
We note that SI units of $I^{\prime}_{kl}$ is $kg \cdot m^2$.

To generalize equation (\ref{1-(1)}) for non-integer 
dimensional space, we should 
represent this equation through 
the dimensionless coordinate variables.
We can introduce the dimensionless values
$x_k = x^{\prime}_k / R_0 , \quad {\bf r}={\bf r}^{\prime}/R_0$, 
where $R_0$ is a characteristic scale, and the density
$\rho ({\bf r},t) = R^{3}_0 \, \rho^{\prime}( {\bf r} \, R_0 ,t)$.
SI units of $\rho$ is $kg$.
We define the following moments of inertia  
$I(t)=R^{-2}_0 I^{\prime}(t)$.
As a result, we obtain 
\be \label{1-(1)-2}
I=\int_{W_3} \rho ({\bf r}) \; {\bf r}^2_{\perp} \; d^3 {\bf r} ,
\ee 
where $x_k$ ($k=1,2,3$) and ${\bf r}$ are dimensionless.
We note that SI units of $I_{kl}$ is $kg$.

This representation allows us to generalize 
equation of the scalar moment of inertia for a fractal material 
\be \label{1-(1)-3}
I^{(D)}(t)=\int_W \rho ({\bf r},t) \; {\bf r}^2_{\perp} \; dV_D ,
\ee 
where $D$ is a mass dimension of fractal material.

\subsection{Moment of inertia of fractal solid ball}

Let us consider a fractal solid ball with radius $R$, and mass $M$.
Note that the component of the radius perpendicular is
\[ {\bf r}^2_{\perp} =(r \ \sin \theta)^2 . \]
Using the integration in a non-integer dimensional 
space, we have 
\be \label{I-D1}
I^{(D)} = \int_{W} d {\bf r}\, (r \ \sin \theta)^2 \rho(r,\theta) 
= \frac{2 \pi^{(D-1)/2}}{\Gamma((D-1)/2)}
\int^{\infty}_0 dr \, r^{D-1} \,
\int^{\pi}_0 d \theta \, sin^{D-2}\theta \,  (r \ \sin \theta)^2 \rho(r,\theta)  .
\ee
For homogeneous materials $\rho({\bf r})=\rho_0$, we get
\[
I^{(D)} = \frac{2 \pi^{(D-1)/2} \, \rho_0}{\Gamma((D-1)/2)}
\int^{\infty}_0 dr \, r^{D+1} \,
\int^{\pi}_0 d \theta \, sin^{D}\theta =
\frac{2 \pi^{(D-1)/2} \, \rho_0}{\Gamma((D-1)/2)} \,
\frac{\pi^{1/2}\, \Gamma (D/2)}{\Gamma(D/2+1)} 
\frac{R^{D+2}}{D+2} = \]
\be \label{I-D2}
=\frac{2 \pi^{D/2} \, \Gamma (D/2) \, \rho_0}{
\Gamma(D/2-1/2) \, \Gamma(D/2+1)} \,
\frac{R^{D+2}}{D+2} =
\frac{\pi^{D/2} \, (D-1) \, \rho_0}{
(D+2) \, \Gamma(D/2+1)} \, R^{D+2} ,
\ee
where we use
\be \int^{\pi}_0 d \theta \, sin^{D}\theta = 
\frac{\pi^{1/2}\, \Gamma (D/2)}{\Gamma(D/2+1)} . \ee
Using the expression for mass (\ref{M-D}), 
we can rewrite (\ref{I-D2}) as
\be \label{I-D4}
I^{(D)} =\frac{2 \, \Gamma (D/2)}{ (D+2) \, 
\Gamma(D/2-1/2)} \, M_D \,  R^{2} 
=\frac{D-1}{D+2} \, M_D \,  R^{2} ,
\ee
where we use $\Gamma (z)=(z-1)\Gamma(z-1)$.
For $D=3$, equation (\ref{I-D4}) gives the well-known 
equation for moment of inertia of non-fractal ball
$I^{3}=(2/5) M R^2$.

\section{Elasticity theory of fractal material}

\subsection{Elasticity theory of non-fractal material}

The linear elastic constitutive relations for isotropic case 
is the well-known Hooke's law that has the form
\be \label{H-0}
\sigma_{ij} = \lambda \varepsilon_{kk} \delta_{ij} + 2 \mu \varepsilon_{ij} ,
\ee
where $\lambda$ and $\mu$ are the Lame coefficients,
$\sigma_{ij}$ is the stress, $\varepsilon_{kl}$ is the strain tensor.
This expression determines the stress tensor in terms of the strain tensor for an isotropic material.

For homogenous and isotropic materials, 
the constitutive relation (\ref{H-0}) gives
the equation for the displacement vector 
fields ${\bf u}={\bf u}({\bf r},t)$ in the form
\be \label{EL-1}
\lambda \, \operatorname{grad} \operatorname{div} {\bf u}
 + 2 \mu \, \Delta \, {\bf u} + {\bf f} = \rho \, D^2_t {\bf u} ,
\ee
where ${\bf f}={\bf u}({\bf r},t)$ is the vector field of external force density.

If the deformation in the material is described by 
${\bf u}({\bf r},t) = u(r,t) \, {\bf e}_r$, then
equation (\ref{EL-1}) has the form
\be \label{EL-2}
(\lambda  + 2 \mu ) \, \Delta \, {\bf u}(r,t) + {\bf f}(r,t) = 
\rho \, D^2_t {\bf u} (r,t) .
\ee

A formal generalization of equations (\ref{EL-2}) 
for fractal material in the framework of
non-integer dimensional models  is
\be \label{F-1}
(\lambda+2 \mu ) \, ^V\Delta^{D}_r \,{\bf u}(r,t) +{\bf f}(r,t) = 
\rho \, D^2_t {\bf u} (r,t) ,
\ee
where $^V\Delta^{D}_r$ is defined by (\ref{V-Delta-D}).
Equation (\ref{F-1}) described dynamics of
displacement vector for fractal elastic materials.

\subsection{Strain and stress in non-integer dimensional space}


Any deformation can be represented as the sum of a pure shear 
and a hydrostatic compression.
To do so for fractal materials, we can use the identity
\be \label{dii-D}
\varepsilon_{kl} = \Bigl( \varepsilon_{kl} - \frac{1}{D} \delta_{kl} \, \varepsilon_{ii} \Bigr) + \frac{1}{D} \delta_{kl} \, \varepsilon_{ii} .
\ee
The first term on the right is a pure shear,
since the sum of diagonal terms is zero.
Here we use the equation $\delta_{ii}=D$ for non-integer dimensional space (for details see Property 4 
in Section 4.3 of \cite{Collins}).
The second term is a hydrostatic compression. 
For $D=3$, equation (\ref{dii-D}) has the well-known form
\be
\varepsilon_{kl} = \Bigl( \varepsilon_{kl} - \frac{1}{3} \delta_{kl} \, \varepsilon_{ii} \Bigr) + \frac{1}{3} \delta_{kl} \, \varepsilon_{ii} ,
\ee
where $\delta_{ii}=3$ is used.

The stress tensor can be represented as
\be
\sigma_{kl} = K \, \varepsilon_{ii} \, \delta_{kl}
+ 2 \mu \, \Bigl( \varepsilon_{kl} - \frac{1}{D} \delta_{kl} \, \varepsilon_{ii} \Bigr) , 
\ee
where the bulk modulus (modulus of hydrostatic compression) 
is related to the Lame coefficients by
\be
K= \lambda + \frac{2 \, \mu}{D} .
\ee

In the hydrostatic compression of a body,
the stress tensor is
\be
\sigma_{kl} = - p \, \delta_{kl} .
\ee
Hence we have 
\be
\sigma_{kk} =- p \, D .
\ee

If we use the Hooke's law for isotropic case in the form, then
\be
\sigma_{ii} = (\lambda \, D + 2\, \mu) \, \varepsilon_{ii}  .
\ee


The components of the strain tensor is 
\be \label{21a}
\varepsilon_{rr} = \frac{\partial u_r}{\partial r} = 
({\bf e}_r,\operatorname{Grad}^{D}_r u_r) . 
\ee
Using (\ref{Div-D}), and the trace of the strain tensor
\be \label{21b}
e (r) = Tr[\varepsilon_{kl}] = \varepsilon_{kk} =
\operatorname{Div}^D_r {\bf u} = 
\frac{\partial u_r}{\partial r} + \frac{D-1}{r} \, u_r .
\ee
Note that we can consider
\[ e(r) - \varepsilon_{rr}(r)=
\operatorname{Div}^D_r {\bf u} - ({\bf e}_r,\operatorname{Grad}^{D}_r u_r) 
= \frac{D-1}{r} \, u_r \]
as a sum of the angle diagonal components
in the spherical coordinates of the strain tensor.
For $D=3$, we have
the well-known sum of the angle diagonal components
in the spherical coordinates of the strain tensor
\[ \varepsilon_{\theta \theta} + 
\varepsilon_{\varphi \varphi} =
\frac{2}{r} \, u_r  . \]
When we consider the fractal medium distributed 
in three-dimensional space 
we can define the effective value of the 
diagonal angular components of the strain tensor
\be \label{Eeff}
\varepsilon^{eff}_{\theta \theta} =
\varepsilon^{eff}_{\varphi \varphi} =
\frac{D-1}{2 r} \, u_r .
\ee


Using (\ref{21a}) and (\ref{21b}), 
the components of the stress tensor $\sigma_{kl}=\sigma_{kl}(r)$ 
in the spherical coordinates is
\be \label{sigma-rr}
\sigma_{rr}(r) = 2 \, \mu \, \varepsilon_{rr}(r) + 
\lambda \, e(r)  =
(2\, \mu + \lambda) \, \frac{\partial u_r}{\partial r} + 
\lambda \, \frac{D-1}{r} \, u_r .
\ee

It is well-known that 
the diagonal angular components of the stress tensor
for $D=3$ in spherical coordinates are
\be
\sigma_{\theta \theta} (r) = 
2 \, \mu \, \varepsilon_{\theta \theta}(r) + 
\lambda \, e(r) , \quad
\sigma_{\varphi \varphi} (r) = 
2 \, \mu \, \varepsilon_{\varphi \varphi}(r) + 
\lambda \, e(r) .
\ee

For the fractal medium distributed in three-dimensional space 
we can define the effective value of 
the diagonal angular components of the stress tensor
\be
\sigma^{eff}_{\theta \theta} (r) = 
2 \, \mu \, \varepsilon^{eff}_{\theta \theta}(r) + 
\lambda \, e(r) ,
\ee
\be
\sigma^{eff}_{\varphi \varphi} (r) = 
2 \, \mu \, \varepsilon^{eff}_{\varphi \varphi}(r) + 
\lambda \, e(r) .
\ee
Using the form of the effective components (\ref{Eeff}), we get
\be \label{sigma-eff}
\sigma^{eff}_{\theta \theta} (r) = 
\sigma^{eff}_{\varphi \varphi} (r) = 
\lambda \, \frac{\partial u_r}{\partial r} + 
(\lambda +  \mu) \, \frac{D-1}{r} \, u_r .
\ee
This equation define the diagonal angular components 
of the stress tensor for spherical coordinates.

\subsection{Equilibrium equation for fractal materials}

For static case, equation (\ref{F-1}), we have
\be \label{F-2}
^V\Delta^{D}_r \, {\bf u}(r) + 
(\lambda+2 \mu )^{-1} \,{\bf f}(r) =  0 ,
\ee
where ${\bf u}= u_r \, {\bf e}_r$ and 
${\bf f}= f (r)\, {\bf e}_r$. 
Here $\lambda$ and $\mu$ are the Lame coefficients.  
Using (\ref{V-Delta-D}), we represent equation (\ref{F-2}) 
in the form
\be \label{F-3}
\frac{\partial^2 u_r(r)}{\partial r^2} +
\frac{D-1}{r} \, \frac{\partial u_r(r)}{\partial r} -
\frac{D-1}{r^2} \, u_r(r)
+ (\lambda+2 \mu )^{-1} \, f (r) =  0.
\ee
The solution of (\ref{F-3}) is
\be
u_r(r) = C_1 \, r + C_2 \, r^{1-D} + I_f (D,r) ,
\ee
where $C_1$ and $C_2$ are constants defined by boundary conditions, and
\be
I_f (D,r) = \frac{r}{D \, (\lambda+2 \mu )} \,
\Bigl( \frac{1}{r^D} \, \int dr \, r^{D} \, f(r)  
-  \int dr \, r \, f(r) \Bigr) .
\ee
For $f(r)=f_0$, we get
\be 
I_f (D,r) = - \frac{f_0 \, r^2}{2 \, (D+1) \, (\lambda+2 \mu)} ,
\ee
and the displacement is
\be
u_r(r) = C_1 \, r + C_2 \, r^{1-D} - \frac{f_0 \, r^2}{(D+1) \, (\lambda+2 \mu)} . 
\ee
The components of the stress tensor $\sigma_{kl}=\sigma_{kl}(r)$ 
for the spherical coordinates can be calculated by equations
(\ref{sigma-rr}) and (\ref{sigma-eff}).


\subsection{Elasticity of fractal hollow ball 
with inside and outside pressures}

Let us determine the deformation of a hollow fractal ball 
with internal radius $R_1$ and external radius $R_2$
with the pressure $p_1$ inside and the pressure $p_2$ outside.

We can use the spherical polar coordinates
with the origin at the center of the ball.
The displacement vector ${\bf u}$ is everywhere radial,
and it is a function of $r=|{\bf r}|$ alone.
Then the equilibrium equation for fractal ball is
\be \label{F-2f0}
(\lambda+2 \mu ) \ ^V\Delta^{D}_r \, {\bf u}(r) =  0 ,
\ee
where ${\bf u}= u_r \, {\bf e}_r$. 
Using (\ref{V-Delta-D}), we represent equation (\ref{F-2}) 
in the form
\be \label{F-3f0}
\frac{\partial^2 u_r(r)}{\partial r^2} +
\frac{D-1}{r} \, \frac{\partial u_r(r)}{\partial r} -
\frac{D-1}{r^2} \, u_r(r) =  0.
\ee
The solution of (\ref{F-3}) is
\be
u_r(r) = C_1 \, r + C_2 \, r^{1-D}  .
\ee
The constants $C_1$ and $C_2$ are determined from  
the boundary conditions for radial stress
\be
\sigma_{rr} (R_1) = - p_1 , \quad \sigma_{rr} (R_2) = - p_2 .
\ee
Using that the radial components of the stress is
\be
\sigma_{rr}(r) = 
(2\, \mu + \lambda) \, \frac{\partial u_r}{\partial r} + 
\lambda \, \frac{D-1}{r} \, u_r  ,
\ee
we get
\be
\sigma_{rr}(r) = (2\, \mu + D\, \lambda) \, C_1 +
2 \, (1-D) \, \mu \, C_2 \, r^{-D} .
\ee
Then we have
\be
(2\, \mu + D\, \lambda) \, C_1 +
2 \, (1-D) \, \mu \, R^{-D}_1 \, C_2  = - p_1 ,
\ee
\be
(2\, \mu + D\, \lambda) \, C_1 +
2 \, (1-D) \, \mu \, R^{-D}_2 \, C_2  = - p_2 .
\ee
As a result, the coefficients have the form
\be
C_1= \frac{- (p_1 \,  R^{-D}_2 - p_2 \,  R^{-D}_1) }{(2\, \mu + D\, \lambda) \, (R^{-D}_2 -R^{-D}_1)} =
 \frac{-(p_2 \,  R^{D}_2 - p_1 \,  R^{D}_1) }{(2\, \mu + D\, \lambda) \, (R^{D}_2 -R^{D}_1)} ,
\ee
\be
C_2=\frac{-  (p_2 - p_1) \, (R_1 \, R_2)^D }{ 2 \, (1-D) \, \mu \, 
(R^{-D}_2 -R^{-D}_1)} 
= \frac{p_2 - p_1}{ 2 \, (1-D) \, \mu \, 
(R^{D}_2 -R^{D}_1)} .
\ee
Then the radial components of the stress is
\be \label{sigma-1}
\sigma_{rr}(r) = 
\frac{-(p_2 \,  R^{D}_2 - p_1 \,  R^{D}_1) }{R^{D}_2 -R^{D}_1} 
+ \frac{(p_2 - p_1) \, (R_1 \, R_2)^D }{R^{D}_2 -R^{D}_1} \, r^{-D} .
\ee
The stress distribution in a ball with pressure $p_1=p$ inside
and $p_2=0$ outside is gives by
\be
\sigma_{rr}(r) = \frac{ p \,  R^{D}_1 }{R^{D}_2 -R^{D}_1} 
- \frac{p}{R^{D}_2 -R^{D}_1} \, r^{-D}
=  \frac{ p \,  R^{D}_1 }{R^{D}_2 -R^{D}_1}
\left( 1 - \left(\frac{R_2}{r}\right)^D \right) .
\ee

The stress distribution in an infinite elastic material with spherical cavity with radius $R$ subjected to hydrostatic compression 
\be
\sigma_{rr}(r) = -p \, \left( 1 - \left(\frac{R}{r}\right)^D \right) .
\ee
that can be obtained by putting $R_1=R$, $R_2 \to \infty$,
$p_1=0$ and $p_2=p$ in equation (\ref{sigma-1}).

\section{Elasticity of fractal material with radial distribution 
in cylinder and pipe}

\subsection{Elasticity of fractal pipe and cylinder}

If we use the cylindrical coordinates with the $z$-axis
and the vector field ${\bf u}({\bf r},t)$ is a purely radial 
\be
{\bf u}=u_r(r) \, {\bf e}_r,
\ee
where ${\bf e}_r={\bf r}/r$, then 
\be \label{Pipe-1}
^V\Delta^D_r {\bf u} = \Bigl(
\frac{\partial^2 u_r(r)}{\partial r^2} 
+ \frac{D-2}{r} \, \frac{\partial u_r}{\partial r} 
- \frac{D-2}{r^2} u_r(r) \Bigr) \, {\bf e}_r.
\ee
Note that we have $(D-2)$ instead of $(D-1)$ in this equation.
For $D=3$, equation (\ref{Pipe-1}) gives 
the well-known equation for elasticity of cylinder and pipe.

If ${\bf u}=u_r(r) \, {\bf e}_r$ is the displacement vector 
for non-fractal materials  in 3-dimensional case ($D=3$), 
then the strain tensor $\varepsilon_{ij}({\bf r})$ 
has the following nonzero components that can be defined by
\be
\varepsilon_{rr} = ({\bf e}_r, \operatorname{grad} u_r) 
= \frac{\partial u_r}{\partial r} , 
\ee
\be
\varepsilon_{\varphi \varphi} = \operatorname{div} {\bf u} -
({\bf e}_r, \operatorname{grad} u_r) = \frac{u_r}{r} ,
\ee
and the invariant
\be
e= \varepsilon_{kk} = \operatorname{div} {\bf u} =
\frac{\partial u_r}{\partial r} +\frac{u_r}{r} .
\ee
These equations with divergence and gradient 
can be generalized for non-integer dimensional case 
for ${\bf u}=u_r(r) \, {\bf e}_r$.

For non-integer dimensional model of fractal materials, 
we can use the definitions for the components of 
displacement vector in non-integer dimensional space
in the form
\be
\varepsilon_{rr} = ({\bf e}_r, \operatorname{Grad}^D_r u_r) ,
\ee
\be
\varepsilon_{\varphi \varphi} = \operatorname{Div}^D_r {\bf u} -
({\bf e}_r, \operatorname{Grad}^D_r u_r) ,
\ee
where we use the invariant
\be
e= \varepsilon_{kk} = \operatorname{Div}^D_r {\bf u} =
\frac{\partial u_r}{\partial r} +\frac{D-2}{r} u_r.
\ee
As a result, we get
\be
\varepsilon_{rr} = \frac{\partial u_r}{\partial r} , 
\ee
\be
\varepsilon_{\varphi \varphi} = \frac{D-2}{r} u_r .
\ee

Let us assume that the elastic constitutive relations 
for fractal material in isotropic case has 
the usual form
\be \label{H-0b}
\sigma_{ij} = \lambda \varepsilon_{kk} \delta_{ij} + 2 \mu \varepsilon_{ij} .
\ee
In this case, the non-zero components of stress tensor are
\be
\sigma_{rr} =2\, \mu \, \varepsilon_{rr} + \lambda \, e =
(2\mu+\lambda) \frac{\partial u_r}{\partial r} +
\lambda \frac{D-2}{r} u_r,
\ee
\be
\sigma_{\varphi \varphi} = 2\mu \, \varepsilon_{\varphi \varphi} + \lambda \, e = 
\lambda \frac{\partial u_r}{\partial r} +
(2\mu+\lambda) \frac{D-2}{r} u_r,
\ee
\be
\sigma_{zz} =2\, \mu \, \varepsilon_{zz} + \lambda \, e = \
\lambda \frac{\partial u_r}{\partial r} +
\lambda \frac{D-2}{r} u_r ,
\ee
where we use $\varepsilon_{zz}=0$.
For $D=3$, we have the usual constitutive relations 
for isotropic case in cylindrical coordinates.

\subsection{Elasticity of cylindrical fractal pipe 
with inside and  outside pressures}

Let us consider the deformation of a fractal solid 
cylindrical pipe with internal radius $R_1$ external radius $R_2$ 
with a inside pressure $p_1$ and outside pressure $p_2$.
We use the cylindrical coordinates with the $z$-axis
along the axis of the pipe. 
When the pressure is uniform along the pipe,
the deformation is a purely radial displacement 
${\bf u}=u_r(r) \, {\bf e}_r$, where ${\bf e}_r={\bf r}/r$.
The equation for the displacement $u_r(r)$ 
in fractal pine is
\be \label{Cyl-1.1}
\frac{\partial^2 u_r(r)}{\partial r^2} + \frac{D-2}{r} \, 
\frac{\partial u_r}{\partial r} - \frac{D-2}{r^2} \, u_r = 0 ,
\ee
where $0<D \le 3$. 
If $D=3$, we get the usual (non-fractal) case.

The general solution of equation (\ref{Cyl-1.1}), 
where $D \ne 1$ and $D \ne 2$, has the form
\be \label{Sol-P-1}
u_r(r) = C_1\, r + C_2\, r^{2-D}  .
\ee
Equations (\ref{Cyl-1.1}) with $D=1$ has the general solution
\be
u_r(r) = C_1\, r + C_2\, r \, \ln (r)  .
\ee
For $D=2$, equations (\ref{Cyl-1.1}) has the solution
\be
u_r(r) = C_1\, + C_2\, r .
\ee
Note that that fractal dimension of the pipe material 
can be $D=1$ or $D=2$. These cases do not correspond 
to the distribution of matter along the line and surface.
These fractal dimensions describe a distribution of matter
in 3-dimensional space (in the volume of pipe) 
such that the mass dimension is equal to $D$.

The constants $C_1$ and $C_2$ are determined by
boundary conditions.
Using that inside pressure is $p_1$ and 
outside pressure is $p_2$, 
we get the boundary condition in the form
\be \label{BCond-1}
\sigma_{rr} (R_1) = - \, p_1 , \quad
\sigma_{rr} (R_2) = - \, p_2 .
\ee
Using (\ref{Sol-P-1}), we get
\be
\frac{\partial u_r}{\partial r} = C_1  + (2-D) \, C_2\, r^{1-D} , \ee
\be
\frac{D-2}{r} u_r = (D-2) \, C_1 + (D-2)\, C_2\, r^{1-D}    .
\ee
Then
\be
\sigma_{rr} =(2\mu+\lambda) \frac{\partial u_r}{\partial r} +
\lambda \frac{D-2}{r} u_r = (2 \mu+ \lambda\, (D-1)) \, C_1 
- 2 \, \mu \, (D-2)\, C_2\, r^{1-D} .
\ee
The boundary condition (\ref{BCond-1}) has the form
\be \label{BCond-1a}
(2 \mu+ \lambda\, (D-1)) \, C_1 
- 2 \, \mu \, (D-2)\, C_2\, R^{1-D}_1  = - \, p_1 , 
\ee
\be \label{BCond-1b}
(2 \mu+ \lambda\, (D-1)) \, C_1 
- 2 \, \mu \, (D-2) \, C_2\, R^{1-D}_2  = - \, p_2 .
\ee
As a result, we have
\be
C_1 = - \frac{ p_1 \, R^{1-D}_2 - p_2 \, R^{1-D}_1 }{ 
(2 \mu+ \lambda\, (D-1)) \, ( R^{1-D}_2 - R^{1-D}_1) } ,
\ee
\be
C_2 = \frac{ p_2  - p_1 }{2 \, \mu \, (D-2) 
\, (R^{1-D}_2 - R^{1-D}_1) } .
\ee
The stress is
\be \label{stress-rr-1}
\sigma_{rr} = - \frac{ p_1 \, R^{1-D}_2 - p_2 \, R^{1-D}_1 }{ 
( R^{1-D}_2 - R^{1-D}_1) }  
- \frac{ p_2  - p_1 }{ (R^{1-D}_2 - R^{1-D}_1) } \, r^{1-D} .
\ee
If $2<D<3$ or $1<D<2$, then we can rewrite 
equation (\ref{stress-rr-1}) in the form
\be \label{stress-rr-2}
\sigma_{rr} = \frac{ p_1 \, R^{D-1}_1 - p_2 \, R^{D-1}_2 }{ 
( R^{D-1}_2 - R^{D-1}_1) }  
- \frac{ p_2  - p_1 }{ (R^{D-1}_2 - R^{D-1}_1) } \, 
\left( \frac{R_1\, R_2}{r} \right)^{D-1} .
\ee
For the boundary conditions 
$\sigma_{rr}(R_2)=0$ and $\sigma_{rr}(R_1)=-p$,
i.e. $p_2=0$ and $p_1=p$ for (\ref{stress-rr-2}),
we have the solution of the form
\be \label{stress-rr-3}
\sigma_{rr} = \frac{ p \, R^{D-1}_1}{ 
( R^{D-1}_2 - R^{D-1}_1) }  \left( 1 -
\left( \frac{R_2}{r} \right)^{D-1} \right) .
\ee
This is the deformation of cylindrical pipe
with a pressure $p$ inside and no pressure outside.
For ($D=3$) we have
\be \label{stress-rr-4}
\sigma_{rr} = \frac{ p \, R^{2}_1}{ 
( R^{2}_2 - R^{2}_1) }  \left( 1 -
\left( \frac{R_2}{r} \right)^{2} \right) 
\ee
that describes the stress of non-fractal material of pipe.

\subsection{Rotating cylindrical fractal pipe}

Let us consider the deformation of a fractal solid
that is describes by equation with external force $f(r)$ 
for the displacement field $u_r(r)$ in fractal pine 
\be \label{Cyl-1.1f}
\frac{\partial^2 u_r(r)}{\partial r^2} + \frac{D-2}{r} \, 
\frac{\partial u_r}{\partial r} - \frac{D-2}{r^2} \, u_r + 
\frac{1}{\lambda +2 \, \mu} f(r) =0 ,
\ee
where $D>0$. 
The general solution of equation (\ref{Cyl-1.1f}) has the form
\be \label{Sol-P-1b}
u_r(r) = C_1\, r + C_2\, r^{2-D}  - 
\frac{1}{(D-1) \, (\lambda +2 \, \mu)} 
\Bigl( \int^{R_2}_{R_1} f (r) \, r \, dr -
r^{2-D} \, \int^{R_2}_{R_1} f (r) \, r^{D-1}\, dr \Bigr).
\ee
Equations (\ref{Cyl-1.1f}) with $D=1$ has the general solution
\be
u_r(r) = C_1\, r + C_2\, r \, \ln (r)  +
\frac{r}{\lambda +2 \, \mu} 
\Bigl( \int^{R_2}_{R_1} f (r) \, \ln (r) \, dr +
\ln(r) \, \int^{R_2}_{R_1} f (r) \, dr \Bigr).
\ee
For $D=2$, equations (\ref{Cyl-1.1f}) has the solution
\be
u_r(r) = C_1\, + C_2\, r  -
\frac{1}{\lambda +2 \, \mu} 
\Bigl( \int^{R_2}_{R_1} f (r) \, r \, dr -
r \, \int^{R_2}_{R_1} f (r) \, dr \Bigr).
\ee

Let us consider the deformation of a fractal solid 
cylindrical pipe with internal radius $R_1$ external radius $R_2$ 
rotating uniformly about its axis with angular velocity $\omega$.
Then the density of the centrifugal force is 
\be
f_r (r) = \rho_0 \, \omega^2 \, r .
\ee
We use the cylindrical coordinates with the $z$-axis
along the axis of the cylinder.
When the pressure is uniform along the pipe,
the deformation is a purely radial displacement 
${\bf u}=u_r(r) \, {\bf e}_r$, where ${\bf e}_r={\bf r}/r$.
The equation for the displacement $u_r(r)$ 
in fractal material is
\be \label{Cyl-2.1}
\frac{\partial^2 u_r(r)}{\partial r^2} + \frac{D-2}{r} \, 
\frac{\partial u_r}{\partial r}  -  \frac{D-2}{r^2} \, u_r =
- \frac{\rho_0 \omega^2}{\lambda + 2 \mu} \, r .
\ee

The general solution of equation (\ref{Cyl-2.1}) has the form
\be \label{Sol-P-2}
u_r(r) = C_1\, r + C_2\, r^{2-D} - A \, r^3 ,
\ee
where 
\be \label{Adef}
A = \frac{\rho_0\, \omega^2}{2(D+1)(\lambda +2\mu)} .
\ee

Using the condition that external forces 
do not act inside and outside the fractal pipe,
we have the boundary condition
\be \label{BCond-2}
\sigma_{rr} (R_1) = 0 , \quad \sigma_{rr} (R_2) = 0 .
\ee

Using (\ref{Sol-P-2}), we get 
\be
\frac{\partial u_r}{\partial r} = C_1  + (2-D) \, C_2\, r^{1-D}  -3 \, A \, r^2 ,
\ee
\be
\frac{D-2}{r} u_r = (D-2) \, C_1 + (D-2)\, C_2\, r^{1-D}  
- A \, (D-2)\, r^2 .
\ee

Then
\[
\sigma_{rr} = (2\mu+\lambda) \frac{\partial u_r}{\partial r} +
\lambda \frac{D-2}{r} u_r = \]
\be
= (2 \mu+ \lambda\, (D-1)) \, C_1 
- 2 \, \mu \, (D-2)\, C_2\, r^{1-D}
- A\, ( 6 \, \mu + \lambda \, (D+1)   )\, r^2 .
\ee
The boundary condition (\ref{BCond-2}) has the form
\be \label{BCond-2a}
(2 \mu+ \lambda\, (D-1)) \, C_1 
- 2 \, \mu \, (D-2)\, C_2\, R^{1-D}_1= 
 A\, ( 6 \, \mu + \lambda \, (D+1)   )\, R^2_1 ,
\ee
\be \label{BCond-2b}
(2 \mu+ \lambda\, (D-1)) \, C_1 
- 2 \, \mu \, (D-2)\, C_2\, R^{1-D}_2= 
 A\, ( 6 \, \mu + \lambda \, (D+1)   )\, R^2_2 .
\ee
Then
\be \label{C1b}
C_1 = \frac{ A\, ( 6 \, \mu + \lambda \, (D+1) )\, 
(R^{D+1}_2 - R^{D+1}_1)}{ (2 \mu+ \lambda\, (D-1)) \, 
( R^{D-1}_2 - R^{D-1}_1)} ,
\ee
\be \label{C2b}
C_2 = \frac{ A\, ( 6 \, \mu + \lambda \, (D+1) )\, 
(R^2_2 - R^2_1) \, (R_1 \, R_2)^{D-1}}{ 2 \, \mu \, (D-2)\,  
( R^{D-1}_2 - R^{D-1}_1)} .
\ee
Substitution of (\ref{C1b}) and (\ref{C2b}) into (\ref{Sol-P-2})
\[ u_r(r) = 
\frac{ A\, ( 6 \, \mu + \lambda \, (D+1) )\, 
(R^{D+1}_2 - R^{D+1}_1)}{ (2 \mu+ \lambda\, (D-1)) \, 
( R^{D-1}_2 - R^{D-1}_1)} \, r + 
\]
\be \label{Sol-P-2b}
+ \frac{ A\, ( 6 \, \mu + \lambda \, (D+1) )\, 
(R^2_2 - R^2_1) \, (R_1 \, R_2)^{D-1}}{ 2 \, \mu \, (D-2)\,  
( R^{D-1}_2 - R^{D-1}_1)}, \, r^{2-D} - A \, r^3 .
\ee
where $A$ is defined by (\ref{Adef}). 

For the fractal cylinder ($R_1=0$, $R_2=R$), we have
\be \label{Sol-P-2c}
u_r(r) = 
\frac{\rho_0\, \omega^2}{2(D+1)(\lambda +2\mu)} \,
\left( 
\frac{ 6 \, \mu + \lambda \, (D+1) }{ 2 \mu+ \lambda\, (D-1)} 
\, R^2 \, r - r^3 \right) .
\ee
For $D=3$, equation (\ref{Sol-P-2c}) gives
\be \label{Sol-P-2e}
u_r(r) = 
\frac{\rho_0\, \omega^2}{8(\lambda +2\mu)} \,
\left( 
\frac{ 3 \, \mu + 2 \, \lambda }{ \mu+ \lambda} 
\, R^2 \, r  -  r^3 \right) .
\ee
Equation (\ref{Sol-P-2e}) describes the displacement field
for elastic cylinder with non-fractal material.

\section{Gradient elasticity model for fractal materials}

\subsection{Gradient elasticity theory of non-fractal material}

In the papers \cite{A1992,AA1992,RA1993} suggested 
to generalize the constitutive relations (\ref{H-0}) by 
the gradient modification that contains the Laplacian in the form
\be \label{H-1}
\sigma_{ij} = \Bigl( \lambda  \varepsilon_{kk} \delta_{ij} + 2 \mu \varepsilon_{ij} \Bigr)
- l^2_s \, \Delta \, \Bigl( \lambda  \varepsilon_{kk} \delta_{ij} + 2 \mu \varepsilon_{ij} \Bigr) ,
\ee
where $l_s$ is the scale parameter \cite{AA2011}.

For homogenous and isotropic materials equation for (\ref{H-1}) has the form
\be \label{EM-3}
\lambda \, \Bigl( 1 \pm l^2_s \Delta \Bigr) \, \operatorname{grad} \operatorname{div} {\bf u}
 + 2 \mu \, \Bigl( 1 \pm l^2_s \Delta \Bigr) \, \Delta \, {\bf u} + {\bf f} = \rho \, D^2_t {\bf u} ,
\ee
where ${\bf f}$ is the vector field of external force density.


Using relation 
\be 
\operatorname{grad} \, \operatorname{div} \, {\bf u} =
\operatorname{curl} \, \operatorname{curl} \, {\bf u} + 
\, \Delta {\bf u} ,
\ee
we can rewrite equation (\ref{EM-3}) as
\be \label{EM-4}
\lambda \, \Bigl( 1 \pm l^2_s \, \Delta \Bigr) \, 
\operatorname{curl} \, \operatorname{curl} \, {\bf u}
 + (\lambda+2 \mu ) \, \Bigl( 1  \pm 
l^2_s \, \Delta \Bigr) \, \Delta {\bf u} + {\bf f} 
= \rho \, D^2_t {\bf u} .
\ee
If we assume that the displacement vector ${\bf u}$
is everywhere radial and it is a function of $r=|{\bf r}|$ 
alone ($u_k=u_k(|{\bf r}|)$), then 
\[ \operatorname{curl} \, {\bf u}=0 . \] 
As a result, equation (\ref{EM-4}) has the form
\be \label{EM-5}
(\lambda+2 \mu ) \, \Bigl( 1 \pm l^2_s \, \Delta \Bigr) 
\, \Delta \,{\bf u} +{\bf f} 
= \rho \, D^2_t {\bf u} .
\ee
This is gradient elasticity equation 
for homogenous and isotropic materials with 
the spherical symmetry.
For the non-gradient model ($l^2_s =0$), equation (\ref{EM-5})
for the displacement gives (\ref{EL-2}).

\subsection{Gradient elasticity of fractal material}

A formal generalization of equations (\ref{EM-5}) 
for fractal material in the framework of continuum models with
non-integer dimensional space, where
the displacement vector ${\bf u}={\bf u}(r,t)$, 
does not depend on the angles,  has the form
\be \label{VL-1}
(\lambda+2 \mu ) \, \Bigl( 1 \pm l^2_s(D) \, ^V\Delta^{D}_r \Bigr) 
\, ^V\Delta^{D}_r \,{\bf u} +{\bf f} = \rho \, D^2_t {\bf u} .
\ee
This is fractional gradient elasticity equation 
for homogenous and isotropic materials with 
the spherical symmetry.
Let us consider equation (\ref{VL-1}) 
for static case ($D^2_t {\bf u}=0$) 
with a minus in front of Laplacian, 
i.e. the GRADELA model for fractal materials \cite{AA2011},
\be \label{VL-2}
(\lambda+2 \mu ) \, \Bigl( 1 - l^2_s(D) \, ^V\Delta^{D}_r \Bigr) 
\, ^V\Delta^{D}_r \,{\bf u} + {\bf f} = 0 .
\ee
We can rewrite this equation as
\be \label{VL-3}
(\, ^V\Delta^{D}_r )^2 \,{\bf u} - 
l^{-2}_s(D) \, ^V\Delta^{D}_r \,{\bf u}  
- (\lambda+2 \mu)^{-1} \, l^{-2}_s(D) \, {\bf f} = 0 .
\ee

Using the vector Laplacian (\ref{V-Delta-D}), we have  
\be \label{VL-4}
^V\Delta^{D}_r {\bf u}(r)=  \Bigl(
\frac{\partial^2 u_r(r)}{\partial r^2} +
\frac{D-1}{r} \, \frac{\partial u_r(r)}{\partial r} 
- \frac{D-1}{r^2} \, u_r(r) \Bigr) \, {\bf e}_r .
\ee
and
\[ (\, ^V\Delta^{D}_r)^2 {\bf u}(r)
=  \Bigl(
\frac{\partial^4 u_r(r)}{\partial r^4} 
+ \frac{2(D-1)}{r} \, \frac{\partial^3 u_r(r)}{\partial r^3} +
\]
\be \label{VL-5}
+ \frac{(D-1)(D-5)}{r^2} \, \frac{\partial^2 u_r(r)}{\partial r^2} 
- \frac{3(D-1)(D-3)}{r^3} \, \frac{\partial u_r(r)}{\partial r}
+ \frac{3(D-1)(D-3)}{r^4} \, u_r(r) \Bigr) \, {\bf e}_r .
\ee
Using (\ref{VL-4} - \ref{VL-5}) and 
${\bf f}(r)=f(r) \, {\bf e}_r$, equation (\ref{VL-3}) 
gives
\[ \frac{\partial^4 u_r(r)}{\partial r^4} 
+ \frac{2(D-1)}{r} \, \frac{\partial^3 u_r(r)}{\partial r^3} + \left(\frac{(D-1)(D-5)}{r^2} - 
l^{-2}_s(D) \right) \, \frac{\partial^2 u_r(r)}{\partial r^2} -
\]
\[
- \left( \frac{3(D-1)(D-3)}{r^3} +  
l^{-2}_s(D) \frac{D-1}{r}\right) 
\, \frac{\partial u_r(r)}{\partial r} +
\]
\be \label{EM-9}
+ \left( \frac{3(D-1)(D-3)}{r^4} +  
l^{-2}_s(D) \frac{D-1}{r^2}\right) \, u_r(r) 
- (\lambda+2 \mu)^{-1} \, l^{-2}_s(D) \, f(r) = 0.
\ee

General solution for the case $f(r)=0$ is
\be
u_r(r) = C_1 \, r + C_2 r^{1-D} - C_3 \, I_I(D,r) -  C_4 \, I_K(D,r) ,
\ee
where $I_I(D,r)$ and $I_K(D,r)$ are the integrals of the Bessel functions
\be
I_I(D,r) = D\, r\, \int dr \, r^{-D-1} \int dr \, r^{D/2+1} I_{D/2}(r/l_s(D)) ,
\ee
\be
I_K(D,r) = D\, r\, \int dr \, r^{-D-1} \int dr \, r^{D/2+1} K_{D/2}(r/l_s(D)) ,
\ee
where $I_{\alpha}(x)$ and $K_{\alpha}(x)$ are Bessel functions of the first and second kinds.

\section{Thermoelasticity of fractal material}

Let us consider a generalization of thermoelasticity 
\cite{TE-1,TE-2,TE-3} for fractal material.
In this section we consider a non-integer-dimensional model
of thermoelasticity of fractal material.
Note that thermoelasticity of fractal materials 
in the framework of the fractional-integral continuum model 
\cite{AP2005-2,PLA2005-1,TarasovSpringer}, 
has been considered by Ostoja-Starzewski in \cite{MOS-TE1,MOS-TE2}.

\subsection{Thermoelastic constitutive relation 
for fractal material}

If the isotropic material is non-uniformly heated, then
the constitutive relation for a thermoelastic material
must include \cite{LL} the term
\be \label{TECR-1}
\sigma_{ij} = \lambda \, \varepsilon_{kk} \, \delta_{ij} + 2 \, \mu \, \varepsilon_{ij} 
- K\, \alpha \, (T-T_0) \, \delta_{ij} ,
\ee
where $\lambda$ and $\mu$ are the Lame coefficients, 
$K$ is the bulk modulus or modulus of compression.
The third term in equation (\ref{TECR-1}) gives 
the additional stresses caused by the change in temperature.

The bulk modulus for fractal materials is related to the Lame coefficients by
\be \label{Klm}
K= \lambda + \frac{2}{D} \mu .
\ee
In this formula, we use the dimension $D$ instead of 3 because 
$\delta_{kk}=D$
for non-integer dimensional space  (see Property 4 in Section 4.3 of \cite{Collins}).

If external forces being absent, then the stress is equal to zero $\sigma_{ij}=0$
and we have a free thermal expansion.
Using $\sigma_{ij}=0$, equation (\ref{TECR-1}) gives
\be
\lambda \varepsilon_{kk} \delta_{ij} + 2 \mu \varepsilon_{ij} - K\, \alpha \, (T-T_0) \, \delta_{ij} = 0 .
\ee
Using $\delta_{kk}=D$ and (\ref{Klm}), we obtain
\be
\varepsilon_{kk} = \alpha \, (T-T_0) .
\ee
Because the function $e=\varepsilon_{kk}$ describes the relative change
of volume caused by deformation, then $\alpha$
is the thermal expansion coefficient of the material \cite{LL}.

The constitutive relation (\ref{TECR-1}) for fractal material 
can be represented in the form
\be \label{TECR-2}
\sigma_{ij} = \lambda \varepsilon_{kk} \delta_{ij} + 2 \mu \varepsilon_{ij} - 
(D\, \lambda +2 \mu)\, \alpha_T \, (T-T_0) \, \delta_{ij} ,
\ee
where we use the dimension $D$ instead of 3 
since $K=(D\, \lambda +2\, \mu)/D$.

\subsection{Thermoelastic equations for fractal material}

For homogenous and isotropic non-fractal materials, 
we have the thermoelasticity equation 
\be \label{TE-1}
\lambda \, \operatorname{grad} \operatorname{div} {\bf u}
 + 2 \mu \, \Delta \, {\bf u} + {\bf f}  
- (3 \lambda +2 \mu)\, \alpha \, \operatorname{grad} T = \rho \, D^2_t {\bf u} ,
\ee
where ${\bf f}$ is the external force density vector field.
For the case ${\bf u}={\bf u}(r,t)$ and ${\bf f}={\bf f}(r,t)$, 
equation (\ref{TE-1}) gives 
\be \label{TE-2}
(\lambda + 2 \mu) \, \Delta \, {\bf u} + {\bf f}  
- (3 \lambda +2 \mu)\, \alpha \, \operatorname{grad} T = \rho \, D^2_t {\bf u} .
\ee

The thermoelasticity equation for fractal materials
in the framework of models with 
non-integer dimensional space has the form
\be \label{TE-3}
(\lambda + 2 \mu) \, ^V\Delta^{D}_r \, {\bf u}(r,t) + {\bf f}(r,t)  
- (D\, \lambda +2 \mu)\, \alpha \, \operatorname{Grad}^D_r T(r,t) = \rho \, D^2_t {\bf u}(r,t) ,
\ee
where we assume ${\bf f}=f_r(r,t) \, {\bf e}_r$
and ${\bf u}=u_r(r,t) \, {\bf e}_r$.
Using (\ref{V-Delta-D}), equation (\ref{TE-3}) 
can be represented in the form
\[
\frac{\partial^2 u_r(r,t)}{\partial r^2} + \frac{D-1}{r} \, 
\frac{\partial u_r(r,t)}{\partial r}  -  \frac{D-1}{r^2} \, u_r (r,t) + \]
\be \label{TE-4}
+ \frac{1}{\lambda + 2 \mu} f_r(r,t)  - \, 
\frac{\alpha \, (D\, \lambda +2 \mu)}{\lambda + 2 \mu}
\frac{\partial T(r,t)}{\partial r} = 
\frac{\rho}{\lambda + 2 \mu} \, D^2_t u_r(r,t) .
\ee
If the fractal material is non-uniformly heated, then
the equation of equilibrium has the form
\be \label{TE-5}
\frac{\partial^2 u_r(r)}{\partial r^2} + \frac{D-1}{r} \, 
\frac{\partial u_r(r)}{\partial r}  -  \frac{D-1}{r^2} \, u_r(r) =
\frac{(D\, \lambda +2 \mu)\, \alpha}{\lambda + 2 \mu} \, 
\frac{\partial T(r)}{\partial r} .
\ee
This is thermoelasticity equation in spherical coordinates 
for pure radial deformation of fractal materials with 
fractal dimension $D$. 
For $D=3$, we get the usual equation for thermoelasticity
of solid ball \cite{LL}. 

General solution of equation (\ref{TE-5}) has the form
\be
u_r(r) = C_1 \, r + \frac{c_2}{r^{D-1}} 
+ 
\frac{(D\, \lambda +2 \mu)\, \alpha}{D \, (\lambda + 2 \mu)}
\Bigl( r \, T(r) - 
\frac{D}{r^{D-1}} \int r^D\, T(r) \, dr \Bigr) ,
\ee
where $C_1$ and $C_2$ are defined by boundary conditions.

\section{Conclusion}

In this paper, continuum models 
with non-integer dimensional space
are suggested to describe isotropic fractal materials.
A generalization of differential vector operators 
for non-integer dimensional space is proposed
to describe elasticity of fractal materials
in the framework of continuum models.
The differential operators of first and second orders 
for non-integer dimensional space
are suggested for rotationally covariant scalar 
and vector functions.
We consider applications for elasticity theory
in the case of spherical and axial symmetries of 
the fractal material.
Elastic properties of 
fractal hollow ball and cylindrical fractal pipe 
with inside and  outside pressures, 
rotating cylindrical fractal elastic pipe are described. 
Equations for thermoelasticity and gradient elasticity 
of fractal materials are solved.

Although the non-integer dimension  
does not reflect all properties 
of the fractal material, 
the suggested models
with non-integer dimensional space
nevertheless allows us to derive 
a number of important conclusions about the behavior
of fractal materials.
Therefore continuum models with non-integer dimensional spaces
can be successfully used to describe elasticity 
and thermoelasticity of fractal materials.

The proposed continuum models of fractal materials
can be extended to more complex fractal materials:
(1) We assume that continuum models
with non-integer dimensional space
can be generalized for anisotropic 
fractal materials \cite{JMP2014};
(2) The non-integer dimensional models 
of fractal elastic materials
can easily be generalized for 
the boundary dimensions $d \ne D-1$ \cite{CNSNS2015}.
(3) We also assume that differential and 
integral operators of fractional orders 
can also be defined for non-integer dimensional spaces
to take into account non-locality of fractal materials.
Note that dimensional continuation of
the Riesz fractional integrals and derivatives \cite{KST}
to generalize differential and integrals of fractional 
orders for non-integer dimensional space 
has been considered in \cite{MA2010}.



\end{document}